\documentclass{article}

\usepackage{graphics}
\usepackage{amsfonts}
\usepackage{amssymb}
\usepackage{amsmath}
\newcommand{\tr}[1]{{\rm tr }\left(#1\right)}   
\newcommand{\h}{{\mathfrak{h}}}

\newcommand{\abs}[1]{\left\vert#1\right\vert}   
\newcommand{\set}[1]{\left\{#1\right\}} 

\newcommand{\ket}[1]{\vert #1\rangle} 
\newcommand{\ketbra}[1]{\vert #1\rangle\langle #1\vert}  
\newcommand{\vketbra}[2]{\vert #1\rangle\langle #2\vert}

\newcommand{\flind}[1]{\mathfrak{L}\kern-8pt{-}(#1)}
\newcommand{\preflind}[1]{\mathfrak{L}_*\kern-12pt{-}\;\;(#1)}

\newcommand{\qds}[2]{{{\mathcal{T}}_{#1}}(#2)} 

\newcommand{\bo}[1]{\mathfrak{L}(#1)}

\newcommand{\1}{\mathbf{1}}

\newcommand{\cs}[1]{\mathfrak{#1}}

\newcommand{\trclass}[1]{\mathfrak{I}_1(#1)}

\newcommand{\C}[1]{{\mathbb C}^{#1}}

\newcommand{\R}[1]{{\mathbb R}^{#1}}

\newcommand{\adj}[1]{{#1}^*}

\newcommand{\ud}[3]{U^{#1\;#2}_{#3}}

\def\bd#1{\mathbf{#1}}

\title{On non-Markovian time evolution in open quantum systems}
\author{Andrzej Kossakowski\\{\footnotesize\it Institute of Physics}\\{\footnotesize\it Nicholaus Copernicus University}\\{\footnotesize\it 87-100 Torun, Poland}\\{\footnotesize\it kossak@fizyka.umk.pl}\\ and \\Rolando Rebolledo \footnote{Partially supported by FONDECYT grant 1030552 and PBCT-ACT13}\\{\footnotesize\it Laboratorio de An\'alisis Estoc\'atico}\\{\footnotesize\it Facultad de
Matem\'aticas}\\{\footnotesize\it Pontificia Universidad Cat\'olica de Chile}\\{\footnotesize\it Casilla 306, Santiago 22,
Chile}\\{\footnotesize\it rrebolle@puc.cl}}
\date{}
\begin{document}
\maketitle
\begin{abstract}
Non-Markovian reduced dynamics of an open system is investigated. In the case the initial state of the reservoir is the vacuum state, an approximation is introduced which makes possible to construct a reduced dynamics which is completely positive.
\end{abstract}

\section{Introduction}
An open system is one coupled to an external environment \cite{Alicki:1987yo,Breuer:2002uq}. Such systems are of fundamental interest since the notion of an isolated system is almost always an idealization and approximation. The interaction between the system and its environment leads to phenomena of decoherence and dissipation, and for this reason recently received intense consideration in quantum information, where decoherence is viewed as a fundamental obstacle to the construction of quantum information processors \cite{Nielsen:2000ly}.

In principle, the von Neumann equation for the total density matrix of the system and the reservoir provides complete theoretical predictions for all the observables. However, this equation is impossible to solve in practice because it takes into account all degrees of freedom of the reservoir. Efforts have been focused on developing direct  methods for the reduced density matrix of the subsystem.

Two well known exact theories of subsystem dynamics are the Feynman-Vernon influence functional theory \cite{Feynman:1963fj,Weiss:1999jo,Legget:1987uf}
and the Nakajima-Zwanzig master equation \cite{Nakajima:1958fx,Zwanzig:1960zh}.

The Feynman-Vernon theory expresses the time evolving reduced density matrix of the subsystem as a path integral over subsystem trajectories weighted by an influence functional which incorporates the effects of the reservoir. In order to calculate the influence functional a path integral over all the reservoir degrees of freedom have to be performed.

The Nakajima-Zwanzig approach employs projection operator techniques to derive an exact equation for the reduced density matrix from the von Neumann equation for the total density matrix. The resulting master equation -an integro-differential equation- is mostly of formal interest since such an exact equation can almost never be solved analytically or even numerically. In contrast, when one makes the Markovian approximation, i.e. when one neglects all reservoir memory effects, the resulting master equation \cite{gorkossud,lindblad76} is formally solvable. Moreover, the required property of complete positivity \cite{Kraus:1983ta} is maintained. A coverted goal of the theory of open quantum systems is a non-Markovian description of time evolution which could at the same time include reservoir memory effects, remaining analytically tractable and retaining complete positivity.

A variety of non-Markovian master equations have been proposed (cf.~[1,\,12\,--\,32]). 
However, the complete positivity of the resulting time evolution is still an important problem to be investigated.

On the other hand, in atomic, molecular and nuclear physics one deals with perturbation of discrete energy levels embedded in continuous spectra. Gamov \cite{Gamov:1928zv} conjectured the existence of eigenvectors of the Hamiltonian corresponding to complex eigenvalues. However, it is impossible within the Hilbert space formulation of Quantum Mechanics because the Hamiltonian --- being a self-adjoint operator --- can have only real spectrum. In the case of simple scattering resonances Bohm and Gadella \cite{Bohm:1989yb} constructed the corresponding Gamov vectors in terms of a suitable extension of quantum theory on the basis of Gel'fand triples (see \cite{Gelfand:1964fk,Maurin:1968kx}). This result was a first step towards a rigorous treatment of irreversibility [37\,--\,45]. 
Physical and mathematical aspects of Gamov states are presented in detail in \cite{Civitarese:2004oq}.

In the present paper an attemp is made to apply these concepts to reduced dynamics of an open system. It is based on the observation that the spectrum of the free Hamiltonian of the system is embedded in the continuous spectrum of the reservoir and as a result of the interaction, the spectrum of the system becomes unstable, which leads to an irreversible evolution of the system.

\section{Reduced Dynamics in the Heisenberg Picture}

Let us consider a finite quantum system $S$ with underlying complex separable Hilbert space $\h^{S}$. The reservoir $R$ will be taken as an infinite quantum system with the Hilbert space $\h^{R}_{\omega}$ determined by the GNS representation $\pi_{\omega}$ induced by the reference state $\omega$, which is assumed to be invariant under the free evolution of $R$.

The composed system $S+R$ is considered to be isolated, i.e., its time evolution is determined by a bounded self-adjoint Hamiltonian $H_{\lambda}$ defined on the space $\h^{S}\otimes\h^R_{\omega}$:
\begin{equation}\label{eq1}
H_{\lambda}\;=\;H^S\otimes\1^R+\1^S\otimes H^R+\lambda V\;=\;H_{0}+\lambda V\,.
\end{equation}
The GNS representation associates to the state $\omega$ a vector in $\h^R_{\omega}$ which we denote by the symbol $\ket{\omega}$. Let $P_{\omega}=\ketbra{\omega}$ be the projection on the state $\omega\in\h^R_{\omega}$. In terms of $P_{\omega}$, we define the following two projectors on the total space $\h^S\otimes\h^R_{\omega}$:
\begin{eqnarray}
P_{0}&=&\1^S\otimes P_{\omega}\\
P_{1}&=&\1-P_{0}\,.
\end{eqnarray}

Notice that the following relation is satisfied:
\begin{equation}\label{2.4}
P_{\alpha}H_{0}\;=\;H_{0}P_{\alpha}\,,
\end{equation}
for $\alpha=0,1$.

Moreover, it is assumed that
\begin{equation}
H^R\ket{\omega}\;=\;0\,.
\end{equation}

The reduced dynamics $\mathcal{T}_t:\bo{\h_S}\to\bo{\h_S}$ and $\mathcal{T}_{*t}:\trclass{\h_S}\to\trclass{\h_S}$ in the Heisenberg and the Schr\"odinger picture, respectively, is defined by the relations:
\begin{eqnarray}\label{2.6}
\langle\varphi,\qds{t}{a}\psi\rangle \;=\; \langle \ud{}{}{t}\varphi\otimes\omega,(a\otimes\1^R)\ud{}{}{t}\psi\otimes\omega\rangle\;
=\;\tr{a\qds{*t}{\vketbra{\psi}{\varphi}}}\,,
\end{eqnarray}
for all $\varphi,\psi\in\h_S$ and any $a\in\bo{\h_S}$, where
\begin{eqnarray}
\ud{}{}{t}&=&\exp (-itH_{\lambda})\,,\\
\ket{\varphi_{t}}&=&U_{t}(\ket{\varphi}\otimes\ket{\omega})\,,\\
\ket{\psi_{t}}&=&U_{t}(\ket{\psi}\otimes\ket{\omega})\,,
\end{eqnarray}
equation \eqref{2.6} can be rewritten in the form
\begin{eqnarray}
\langle \varphi,\qds{t}{a}\psi\rangle&=&\langle\varphi_{t}, (a\otimes\1^R)\psi_{t}\rangle\nonumber\\
&=&\langle P_{0}\varphi_{t},(a\otimes\1^R)P_{0}\psi_{t}\rangle+\langle P_{1}\varphi_{t},(a\otimes\1^{R})P_{1}\psi_{t}\rangle\,.\label{2.11}
\end{eqnarray}
It follows from the above relation that the reduced dynamics is completely determined by the solution of the Schr\"odinger equation
\begin{equation}\label{2.12}
\frac{d\ket{\varphi_{t}}}{dt}\;=\;-iH_{\lambda}\ket{\varphi_{t}}\,,
\end{equation}
with the initial condition
\begin{equation}\label{2.13}
\lim_{t\to 0}\ket{\varphi_{t}}\;=\;\ket{\varphi}\otimes\ket{\omega}\,.
\end{equation}
Using the projectors $P_{0}$, $P_{1}$ and \eqref{2.4} one finds that \eqref{2.12} and \eqref{2.13} are equivalent to the following system of differential equations
\begin{eqnarray}
\frac{dP_{0}\ket{\varphi_{t}}}{dt}&=&iP_{0}H_{\lambda}P_{0}P_{0}\ket{\varphi_{t}}-i\lambda P_{0}VP_{1}P_{1}\ket{\varphi_{t}}\,,\label{2.14}\\
\frac{dP_{1}\ket{\varphi_{t}}}{dt}&=&-i\lambda P_{1}VP_{0}P_{0}\ket{\varphi_{t}}-P_{1}H_{\lambda}P_{1}P_{1}\ket{\varphi_{t}},\label{2.15}
\end{eqnarray}
with the initial conditions
\begin{eqnarray}
\lim_{t\to 0}P_{0}\ket{\varphi_{t}}&=&\ket{\varphi}\otimes\ket{\omega}\,,\label{2.16}\\
\lim_{t\to 0}P_{1}\ket{\varphi_{t}}&=&0\,.\label{2.17}
\end{eqnarray}
 
The ``variation of constants'' method applied to \eqref{2.15} with \eqref{2.17} gives $P_{1}\ket{\varphi_{t}}$ in terms of $P_{0}\ket{\varphi_{t}}$:
\begin{equation}\label{2.18}
P_{1}\ket{\varphi_{t}}\;=\;-i\lambda\int\limits_{0}^t\left( e^{-i(t-s)P_{1}H_{\lambda}P_{1}}P_{1}VP_{0}P_{0}\ket{\varphi_{s}}\right)ds,
\end{equation}
which inserted in \eqref{2.14} yields
\begin{equation}\label{2.19}
\frac{dP_{0}\ket{\varphi_{t}}}{dt}\;=\;-iP_{0}H_{\lambda}P_{0}P_{0}\ket{\varphi_{t}}-\lambda^2
\int\limits_{0}^t\left(P_{0}VP_{1}e^{-i(t-s)P_{1}H_{\lambda}P_{1}}P_{1}VP_{0}P_{0}\ket{\varphi_{s}}\right)ds\,.
\end{equation}
$P_{0}\ket{\varphi_{t}}$ is first obtained solving \eqref{2.19} under the initial condition \eqref{2.16}. After that, $P_{1}\ket{\varphi_{t}}$ follows from \eqref{2.18} and consequently, the reduced dynamics \eqref{2.11} is determined. This dynamics is completely positive by definition.

Notice that $P_{0}\ket{\varphi_{t}}$ is determined by
\begin{equation}\label{2.20}
P_{0}U_{t}P_{0}\;=\;P_{0}e^{-itH_{\lambda}}P_{0}\,,
\end{equation}
and \eqref{2.19} can be rewritten in the form
\begin{equation}\label{2.21}
\frac{dP_{0}U_{t}P_{0}}{dt}=-iP_{0}H_{\lambda}P_{0}P_{0}U_{t}P_{0}-\lambda^2
\int\limits_{0}^t\left(P_{0}VP_{1}e^{-i(t-s)P_{1}H_{\lambda}P_{1}}P_{1}VP_{0}P_{0}U_{s}P_{0}\right)ds\,.
\end{equation}

Let denote $P_{0}U(p)P_{0}$ the Laplace transform of $P_{0}U_{t}P_{0}$, that is,
\begin{equation}\label{2.22}
P_{0}U(p)P_{0}\;=\;P_{0}(p+iH_{\lambda})^{-1}P_{0}\,.
\end{equation}
This means that $P_{0}U(p)P_{0}$  is the reduced resolvent. On the other hand, taking the Laplace transform of \eqref{2.21} and using \eqref{2.22} it follows that
\begin{equation}\label{2.23}
P_{0}(p+iH_{\lambda})^{-1}P_{0}\;=\;
\left[p+iP_{0}H_{\lambda}P_{0}+\lambda^2P_{0}VP_{1}
(p+iP_{1}VP_{1})^{-1}P_{1}VP_{0}\right]^{-1}P_{0}\,,
\end{equation}
which can also be derived via the resolvent equation.

It is clear that the properties of $P_{0}\ket{\varphi_{t}}$ and $P_{1}\ket{\varphi_{t}}$ are determined by analytical properties of the reduced resolvent.

\section{The Friedrichs Approximation}

It is easy to understand that the formalism developed so far can not be applied in practice due to the complicated nature of the unitary operator $\exp (-iP_{1}H_{\lambda}P_{1})$ which appears in \eqref{2.18}, \eqref{2.19} and \eqref{2.21}. In fact, the memory terms in \eqref{2.18}, contain an infinite number of multi-time correlation functions. 

It is worth noticing that the expression becomes simpler as soon as one assumes the additional condition
\begin{equation}\label{3.1}
P_{1}VP_{1}\;=\;0\,.
\end{equation}
 
Condition \eqref{3.1} will be called the {\em Friedrichs condition} since it is satisfied in the Friedrichs model \cite{Friedrichs:1948uq} as well as in its $N$-level versions (see [48\,--\,52]). 
The simplification is due to the fact that in this case,
\begin{equation}\label{3.2}
P_{1}H_{\lambda}P_{1}\;=\;P_{1}H_{0}P_{1}\;=\;P_{1}H_{0}\,.
\end{equation}

The meaning of condition \eqref{3.1} is that only one subspace of $\h^R_{\omega}$ plays an essential role.

In general the interaction Hamiltonian $V$ does not satisfy \eqref{3.1}, but $V$ can always be decomposed as follows
\begin{equation}\label{3.3}
V\;=\;V-P_{1}VP_{1}+P_{1}VP_{1}\;=\;\widetilde{V}+P_{1}VP_{1}\,,
\end{equation}
where $\widetilde{V}=V-P_{1}VP_{1}=P_{0}V+VP_{0}-P_{0}VP_{0}$. This operator has the following properties:
\begin{eqnarray}
P_{1}\widetilde{V}P_{1}&=&0\,,\label{3.5}\\
P_{1}\widetilde{V}P_{0}&=&P_{1}VP_{0}\,,\label{3.6}\\
P_{0}\widetilde{V}P_{1}&=&P_{0}VP_{1}\,.\label{3.7}
\end{eqnarray}
So that the Friedrichs approximation consists of replacing the interaction Hamiltonian $V$ by $\widetilde{V}$.

In the case $\h^S=\C{n}$, the interaction Hamiltonian may be written in the form:
\begin{equation}\label{3.8}
V=\sum_{i,j=1}^n\vketbra{e_{i}}{e_{j}}\otimes F_{i,j},
\end{equation}
where $\ket{e_{1}},\ldots,\ket{e_{n}}$ is an orthonormal basis in $\C{n}$ and the operators $F_{i,j}$ satisfy $\adj{F_{i,j}}=F_{j,i}$. The operator $\widetilde{V}$ is then expressed by
\begin{equation}\label{3.10}
\widetilde{V}=\sum_{i,j=1}^n\vketbra{e_{i}}{e_{j}}\otimes\Big(F_{i,j}\ketbra{\omega}+\vketbra{\omega}{F_{i,j}\omega}-\langle\omega,F_{i,j}\omega\rangle\ketbra{\omega}\Big)\,,
\end{equation}
that means that the only transitions in the reservoir are
\[\ket{\omega}\to\ket{\omega}\,,\qquad\ket{\omega}\to F_{i,j}\ket{\omega}\,,\qquad F_{i,j}\ket{\omega}\to \ket{\omega}\,.
\]

To illustrate how the Friedrichs approximation works, let us consider an $n$-level system interacting with an electromagnetic field in the vacuum state. The corresponding Hilbert space for $S$ is $\h^S=\C{n}$ and the Hamiltonian $H^S$ can be written in the form
\begin{equation}\label{3.12}
H^S\;=\;\sum_{i=1}^n\epsilon_{i}\vketbra{e_{i}}{e_{i}}\,.
\end{equation}

For simplicity, it is assumed that the spectrum of $H^S$ is non-degenerate.

The reservoir is chosen to consists of the quantized electromagnetic field. The modes of the electromagnetic field are indexed by $k=(\bd{k},\lambda)$, where $\bd{k}\in\R{3}$, $\lambda\in Z_{2}=\set{1,2}$. The Hilbert space of one photon states is $\h_{1}=L^{2}(\R{3}\otimes Z_{2})$, and for any element $f\in\h_{1}$ we write
\[\int f(k)dk\;=\;\sum_{\lambda=1,2}\int_{\R{3}}f(\bd{k},\lambda)d^3\bd{k}\,.
\]

At zero temperature, the Hilbert space of pure states of the reservoir is the Fock space $\cs{F}=\bigoplus_{k=0}^\infty\h_{k}$, where $\h_{0}=\C{}$ and $\h_{k}$ is the symmetrized $k$-fold tensor product of $\h_{1}$. In $\cs{F}$ the creation and annihilation operators are introduced in the standard manner.
In particular,
\begin{equation}\label{3.14}
a(f)\;=\;\int a(k)f(k)dk\,,
\end{equation}
for any $f\in\h_{1}$ and the vacuum state is determined by the condition
\begin{equation}\label{3.15}
a(f)\ket{\omega}\;=\;0\,,
\end{equation}
for all $f\in\h_{1}$.

The free evolution of the reservoir is defined by the formal Hamiltonian
\begin{equation}\label{3.16}
H^R=\int \abs{\bd{k}}a^*(k)a(k)dk\,.
\end{equation}

The simplest interaction Hamiltonian $V$ can be chosen in the form
\begin{equation}\label{3.17}
V\;=\;
\sum_{i,j}^n\Big(\vketbra{e_{i}}{e_{j}}\otimes a^*(f_{i,j})+\vketbra{e_{j}}{e_{i}}\otimes a(f_{i,j})\Big)\,,
\end{equation}
where $f_{i,j}\in\h_{1}$.

Making the Friedrichs approximation, 
i.e.~$V\mapsto \widetilde{V}=P_{0}V+VP_{0}-P_{0}VP_{0}$ with
\[
P_{0}\;=\;\1^S\otimes\ketbra{\omega}\,,
\]
one obtains
\begin{equation}\label{3.20}
\widetilde{V}\;=\;
\sum_{i,j=1}^n\Big(\vketbra{e_{i}}{e_{j}}\otimes a^*(f_{i,j})\ketbra{\omega}+
\vketbra{e_{j}}{e_{i}}\otimes \ketbra{\omega}a(f_{i,j})\Big)\,.
\end{equation}

Now, applying \eqref{2.21} or \eqref{2.23}, one finds
\begin{equation}\label{3.21}
P_{0}(p+iH_{\lambda})^{-1}P_{0}\;=\;G^{-1}(p)\otimes\ketbra{\omega}\,,
\end{equation}
where
\begin{equation}\label{3.22}
\langle e_{k},G(p)e_{\ell}\rangle\;=\;
\delta_{k,\ell}(p+i\epsilon_{k})+\lambda^2\sum_{m=1}^n\sum_{\lambda=1,2}
\int_{\R{3}}\frac{f_{m,k}(\bd{k},\lambda)\overline{f_{m,\ell}}(\bd{k},\lambda)}
{p+i\epsilon_{m}+i\abs{\bd{k}}}d^3\bd{k}\,.
\end{equation}
The properties of $G(p)$ depend on the choice of the functions $f_{k,\ell}(\bd{k},\lambda)$. 
The above model is an analogue of the one due to Friedrichs.

It should be pointed out that the solvability of the above model is related to the invariance of the subspace $\C{n}\otimes\omega\oplus\C{n}\otimes\h_{1}$ under the action of $H_{0}+\widetilde{V}$. In spin-boson model (c.f.~\cite{Breuer:2002uq}) as well as models considered in [52\,--\,57], 
the existence of such an invariant subspace is due to an additional constant of motion.

\section{The Friedrich Approximation and non-Markovian Master Equation}

From \eqref{2.11} it follows that the reduced dynamics in the Heisenberg picture can also be written in the form
\begin{equation}\label{4.1}
\qds{t}{a}\otimes P_{\omega}\;=\;P_{0}e^{itH}(a\otimes\1^R)e^{-itH}P_{0}\;=\;P_{0}a(t)P_{0}\,,
\end{equation}
where $a(t)$ is the solution of the Heisenberg equation
\begin{equation}\label{4.2}
\frac{da (t) }{dt}\;=\;i[H,a(t)]\,,
\end{equation}
with the initial condition
\begin{equation}\label{4.3}
\lim_{t\to 0}a (t)\;=\;a\otimes\1^R\,.
\end{equation}
It is interesting to find out the master equation for $\qds{t}{a}\otimes P_{\omega}$ under the hypothesis that the interaction Hamiltonian satisfies the Friedrichs condition of the previous section, that is $V$ has the form $V=P_{0}VP_{0}+P_{0}VP_{1}+P_{1}VP_{0}$. For any $X\in\bo{\cs{H}}$, we denote $X_{\alpha\beta}=P_{\alpha}XP_{\beta}$ for $\alpha,\beta=0,1$. So that $V=V_{00}+V_{01}+V_{10}$.

Thus, \eqref{4.2} becomes equivalent to the following system of equations:
\begin{eqnarray}
\frac{da_{00}(t)}{dt}&=&i[H_{00},a_{00}(t)]+i(V_{01}a_{10}(t)-a_{01}(t)V_{10})\,,\\
\frac{da_{01}(t)}{dt}&=&i(V_{01}a_{11}(t)-a_{00}(t)V_{01})\,,\\
\frac{da_{10}(t)}{dt}&=&i(V_{10}a_{00}(t)-a_{11}(t)V_{10})\,,\\
\frac{da_{11}(t)}{dt}&=&i[H^0_{11},a_{11}(t)]+i(V_{10}a_{01}(t)-a_{10}(t)V_{01})\,,
\end{eqnarray}
with the initial conditions
\begin{eqnarray}
\lim_{t\to 0}a_{00}(t)&=&a\otimes P_{\omega}\,,\\
\lim_{t\to 0}a_{01}(t)&=& \lim_{t\to 0}a_{10}(t)\;=\;\lim_{t\to 0}a_{11}(t)\;=\;0\,.
\end{eqnarray}
Eliminating $a_{10}(t)$ and $a_{01}(t)$ in the previous system yields
\begin{eqnarray}
\frac{d^2a_{00}(t)}{dt^2}
&=& i\Big[H_{00},\frac{da_{00}(t)}{dt}\Big]+2V_{01}a_{11}(t)V_{10}-
\set{a_{00}(t),V_{01}V_{10}}\,,\\
\frac{d^2a_{11}(t)}{dt^2}&=&i\Big[H^0_{11},\frac{da_{11}(t)}{dt}\Big]+
2V_{10}a_{00}(t)V_{01}-\set{a_{11}(t),V_{10}V_{01}}\,,
\end{eqnarray}
with the initial copnditions
\begin{eqnarray}
\lim_{t\to 0}a_{00}(t)&=&a\otimes P_{\omega}\,,\\
\lim_{t\to 0}\frac{da_{00}(t)}{dt}&=&i[H_{00},a\otimes P_{\omega}]\,,\\
\lim_{t\to 0}a_{11}(t)&=&0\,,\\
\lim_{t\to 0}\frac{da_{11}(t)}{dt}&=&0\,.
\end{eqnarray}

Finally, using the following matrix notations:
\[\mathbf{A} (t)=   \left(\!\!\begin{array}{cc} 
      a_{00}(t)& 0\\
      0 & a_{11}(t) \\
   \end{array}\!\!\right),\quad\mathbf{H}=  \left(\!\! 
\begin{array}{cc}          H_{00} & 0 \\
         0 & H^0_{11} \\
      \end{array}\!\!\right),\quad\mathbf{V}=   \left(\!\!
\begin{array}{cc}             0 & V_{01}\\
            V_{10}& 0 \\
         \end{array}\!\!\right)={\mathbf V}^*,
\]
the above system of equations can be expressed in the form:
\begin{equation}\label{4.16}
\frac{d^2\mathbf{A}(t)}{dt^2}\;=\;
i[\mathbf{H},\frac{d\mathbf{A}(t)}{dt}]-\lambda^2[\mathbf{V},[\mathbf{V},\mathbf{A}(t)]]\,.
\end{equation}

Equation \eqref{4.16} is local in time but rather untractable.

\section*{Acknowledgements}

R.\,R. acknowledges a partiall support by FONDECYT grant 1030552 and PBCT-ACT13.

\end{document}